\documentclass[twocolumn,prepintnumbers,showpacs,superscriptaddress,amsmath,amssymb,prb, floatfix]{revtex4}

\usepackage{graphicx}
\usepackage{dcolumn}
\usepackage{bm}
\usepackage{hyperref}
\usepackage{verbatim}
\usepackage{color}

\begin{document}

\title{Spin--orbital interaction for face-sharing octahedra:
Realization of a highly symmetric SU(4) model}

\author{K.~I.~Kugel} \affiliation{Institute for Theoretical and
Applied Electrodynamics, Russian Academy of Sciences, Izhorskaya
str. 13, 125412 Moscow, Russia}

\author{D.~I.~Khomskii} \affiliation{$II.$ Physikalisches
Institut, Universit\"at zu K\"oln, Z\"ulpicher Str. 77, 50937
K\"oln, Germany}

\author{A.~O.~Sboychakov} \affiliation{Institute for Theoretical
and Applied Electrodynamics, Russian Academy of Sciences,
Izhorskaya str. 13, 125412 Moscow, Russia}

\author{S.~V.~Streltsov} \affiliation{Institute of Metal Physics,
Ural Branch, Russian Academy of Sciences, S. Kovalevskaya Str. 18,
Ekaterinburg, 620990 Russia} \affiliation{Ural Federal University,
Mira Str.  19, Ekaterinburg, 620002 Russia}

\begin{abstract} Specific features of orbital and spin structure
of transition metal compounds in the case of the face-sharing
MO$_6$ octahedra are analyzed. In this geometry, we consider the
form of the spin--orbital Hamiltonian for transition metal ions
with double ($e_g^{\sigma}$) or triple ($t_{2g}$) orbital
degeneracy. Trigonal distortions typical of the structures with
face-sharing octahedra lead to splitting of $t_{2g}$ orbitals into
an $a_{1g}$ singlet and $e_g^{\pi}$ doublet. For both doublets
($e_g^{\sigma}$ and $e_g^{\pi}$), in the case of one electron or
hole per site, we arrive at a symmetric model with the orbital and
spin interaction of the Heisenberg type and the Hamiltonian of
unexpectedly high symmetry: SU(4). Thus, many real materials with
this geometry can serve as a testing ground for checking the
prediction of this interesting theoretical  model.  We also
compare general trends in spin--orbital (``Kugel--Khomskii")
exchange interaction for three typical situations: those of MO$_6$
octahedra with common corner, common edge, and the present case of
common face, which has not been considered yet. \end{abstract}

\pacs{75.25.Dk, 
75.30.Et, 
75.47.Lx, 
71.27.+a, 
71.70.Ej, 
75.10.Dg  
}

\date{\today}

\maketitle


\section{Introduction}

Systems with orbital degeneracy usually exhibit quite diverse properties, often much different from those of purely spin systems.~\cite{Gooden_book,Khomskii_book2014,OlesJPCM2012} In
particular, the coupling between orbital and spin degrees of freedom, besides being of practical importance for many specific materials, leads to several interesting theoretical models, such as spin--orbital model (often called the Kugel--Khomskii model)~\cite{KK-JETP,KugKhoUFN82}, the popular nowadays compass
model~\cite{KugKhoUFN82,Nussinov}, a particular version of which
is the renowned Kitaev model.~\cite{Kitaev,Nussinov}.

It turns out that the specific features of one or another system with spin and orbital degeneracy strongly depend on the local geometry. The most typical cases, widely discussed in the literature, are those with MO$_6$ octahedra (M is a transition metal ion) sharing common oxygen (or common corner), typical e.g. of perovskites like LaMnO$_3$ or layered systems such as La$_2$CuO$_4$, and the situation with two common oxygens for neighboring octahedra (octahedra with common edge), met in many layered systems with triangular lattices such as NaCoO$_2$ and LiNiO$_2$. The features of spin--orbital systems in both these cases were studied in detail, see e.g. Refs.~\onlinecite{Khomskii_book2014,KhomskiiPhScr2005}. However, there exists yet the third typical geometry, which is also very often met in many real materials -- the case of octahedra with common face (three common oxygens). Strangely enough, this case has not been actually considered in the literature. To fill this gap and to develop a theoretical description of spin--orbital (Kugel--Khomskii) model for this ``third case"  is the main goal of the present paper.

Interestingly enough, after fulfilling this program, we have found out that the resulting model has a very symmetric form -- more symmetric that for the cases of common corner or common edge. The resulting Hamiltonian in the main approximation turned out to have a very high symmetry: SU(4). Actually, SU(4) model appeared already in the very first treatment of these models~\cite{KK-JETP,KugKhoUFN82}  for the ``artificial" illustrative case, in which for doubly-degenerate orbitals only the diagonal inter-site hopping exists, that is,
\begin{equation} \label{hopping-condition}
t_{11} = t_{22} = t,\quad t_{12} = 0,
\end{equation}
where 1 and 2 are the indices denoting two degenerate orbitals. The resulting exchange Hamiltonian, derived from the degenerate Hubbard model in the strong coupling limit $t/U \ll 1$ ($U$ is the on-site Coulomb repulsion), written in terms of spin $s$=1/2 and pseudospin $\tau =1/2$ operators describing doubly-degenerate orbitals, has a very symmetric form
 \begin{eqnarray}
 H &=&\frac{t^2}{U}\sum_{\langle i,j\rangle }\left(\frac{1}{2}
 + 2{\bm s}_i{\bm s}_j\right)
\left(\frac{1}{2} + 2{\bm \tau}_i{\bm \tau}_j\right).
\label{KK_symmetric}
 \end{eqnarray}
This Hamiltonian not only has SU(2)$\times$SU(2) symmetry (it contains scalar products of $\bm s$ and $\bm \tau$ vector operators), but it shows even much higher SU(4) symmetry (interchange of 4 possible states: 1$\uparrow$, 1$\downarrow$, 2$\uparrow$, 2$\downarrow$).

The SU(4) spin--orbital model was extensively discussed in the literature with the main emphasis on novel quantum states (exact solution of the 1D model~\cite{Sutherland}; the presence of three Goldstone modes~\cite{Sutherland,Frischmuth}; the gap formation~\cite{ZhangPRB2003}; spin-orbital singlets on plaquettes in square lattice~\cite{Li} and in two-leg ladders~\cite{vdBosschePRL2001}; spontaneous symmetry breaking with the formation of dimer columns~\cite{CorbozPRL2011}; real spin-orbital liquid on honeycomb lattice~\cite{CorbozPRX2012}). There were also some attempts to apply this model to real materials.~\cite{Pati1998,PencPRB2003,VernayPRB2004,MilaPRB2014} Recently the SU(4) model (or more general SU(N) model with N ``colors") has been applied also to cold atoms on a lattice.~\cite{GorshkovNatPhys2010} However, especially as to real transition metal compounds, these applications were still rather questionable.~\cite{MostKhomPRL2002,KhomMostJPhA2003} In the present paper, we demonstrate that there exists situation in transition metal solids, in which the SU(4) physics might be close to reality -- this is the case of spin--orbital systems with face-sharing MO$_6$ octahedra. If we include the terms in the effective exchange Hamiltonian, which break this SU(4) symmetry, see Appendix C, such terms are usually much weaker than the dominant SU(4) exchange, so that in any case the SU(4) physics would dominate the properties of a system in a broad temperature interval $J'<T<J$, where $J$ is the scale of SU(4) terms in the exchange, and $J'$ -– that of symmetry-breaking terms (typically in systems with 3$d$ elements $J' \sim 0.1J$). Even at $T=0$ strong  quantum fluctuations in SU(4) model, especially in one-dimensional systems, may overcome the effect of symmetry-breaking terms.
\begin{figure}[t!] \centering
\includegraphics[width=0.35\columnwidth]{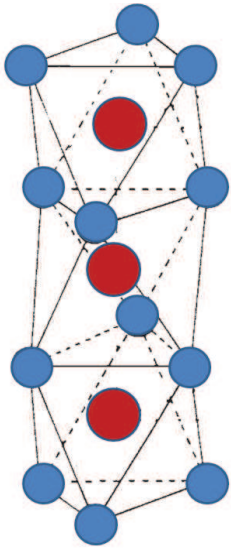}
\caption{(Color online) A chain of face-sharing octahedra. Large (red) and small (blue) circles denote metal and ligand ions, respectively.}\label{face_sharing}
\end{figure}

As far as the actual materials are concerned, in most typical and best studied geometries, such as in systems like perovskites, with
corner-sharing MO$_6$ octahedra and with $\sim 180^{\circ}$ M--O--M bonds (M is the transition metal), the problem is that
conditions~\eqref{hopping-condition} required for SU(4) model~\eqref{KK_symmetric} are not fulfilled. In effect, whereas
the spin part of the spin--orbital exchange is of the Heisenberg ${\bm s}_i{\bm s}_j$ type [SU(2)], the orbital part of the exchange turns out to be very anisotropic, containing terms of the type $\tau^z \tau^z$, $\tau^x \tau^x$, $\tau^z \tau^x$, and also some linear terms, but not, for example $\tau^y \tau^y$. (The latter terms can appear for complex combinations of the basis orbitals, which usually do not lead to static lattice distortions but may be sometime important giving rise to quite exotic types of the ground state.~\cite{vdBrinkKhomPRB2001}) Also for another well-studied case, that with edge-sharing octahedra and with 90$^{\circ}$ M--O--M bonds, the situation is more complicated: sometimes the orbital part of the exchange is anisotropic, and in some cases the leading term in the exchange, $\sim t^2/U$, drops out at all and the remaining exchange depends on the Hund's rule exchange (not included above).~\cite{Gooden_book,MostKhomPRL2002,StrKhomPRB2008pyroxenes}

The third, much less studied situation, that with the face-sharing MO$_6$  octahedra (see Fig.~\ref{face_sharing}) is considered below. In situation with face-sharing octahedra, one naturally obtains for the doubly-degenerate case ($e_g^{\sigma}$ orbitals, or $e_g^{\pi}$ orbitals obtained from triply-degenerate $t_{2g}$ orbitals due to trigonal crystal field, typical for this geometry) that the resulting spin--orbital (Kugel--Khomskii) model is of the type of Eq.~\eqref{KK_symmetric}, i.e. it is the SU(4)-symmetric model. Thus, the systems with this geometry, which are in fact quite abundant among transition metal compounds, represent an actual realization of the high-symmetry SU(4) model, and can provide a natural testing ground for it. The experimental study of the systems with face-sharing arrays may thus allow for verification of the predictions of this model, such as the strong spin--orbital entanglement, and the presence of three Goldstone modes.

Experimentally, there are many transition metal compounds with the face-sharing geometry. Such materials include for example hexagonal crystals like BaCoO$_3$~\cite{YamauraJSStCh1999BaCoO3}, BaVS$_3$~\cite{Gardner1969} or CsCuCl$_3$~\cite{HirotsuJPhC1977CsCuCl3}, containing infinite columns of face-sharing ML$_6$ octahedra (L stands here for ligands O, Cl, S, ...), as shown in Fig.~\ref{face_sharing}. Many other similar systems have finite face-sharing blocks, e.g. BaIrO$_3$,~\cite{SiegristJLCM1991BaIrO3}
BaRuO$_3$~\cite{HongJSStCh1997BaRuO3,ZhaoJSStCh2007BaRuO3}, or Ba$_4$Ru$_3$O$_{10}$~\cite{CarimJSStCh2000Ba4Ru3O10,Klein2011}
with blocks of three such face-sharing octahedra, connected between themselves by common corners; or blocks of two such
octahedra as in large series of systems with general formula A$_3$(M1)(M2)$_2$O$_9$~\cite{KohlReinenZanorgAlChem1977Ba3CuSb2O9,Kimber2012,Senn2013,Senn2013a,Fernandez1980,Rijssenbeek1999,Rijssenbeek1999a,Streltsov2013a,VonTreiberU.1982}, where $A$ is Ba, Ca, Sr, Li, or Na, and face-sharing M2O$_6$ octahedra of transition metals are separated by M1O$_6$ octahedra (which have common corners with M2O$_6$). Such systems have very diverse properties: some of them are metallic~\cite{Rijssenbeek1999a}, others are insulators~\cite{Klein2011} or undergo metal--insulator transition~\cite{Kimber2012}; despite similar crystal structures they may have charge ordered~\cite{Kimber2012} or uniform~\cite{Senn2013a} charge states and their magnetic properties are also quite different changing from the singlet ground state~\cite{Senn2013,Fernandez1980}, to the situations when part of the magnetic moments turn out be suppressed~\cite{Klein2011,Streltsov2012a} and to ferro- or antiferromagnetic order~\cite{YamauraJSStCh1999BaCoO3,Rijssenbeek1999,Lightfoot1990}. However, in any case, the first problem to consider for such systems is that of a possible orbital and magnetic exchange in this geometry. The analysis of this problem is the main task of the present paper.

In Section II, we formulate a minimal model for the face-sharing geometry, which is in fact the Hubbard model taking into account the orbital degrees of freedom. In Sections III and IV, we consider the chains of face-sharing octahedra with $e_g$ and $t_{2g}$, respectively, and demonstrate that in both cases we arrive at a highly symmetrical spin--orbital model. The obtained results are discussed in Section V. More technical issues are discussed in appendices. In Appendix A, we show that trigonal distortions characteristic of the face-sharing geometry do not affect the symmetric form of the effective spin--orbital Hamiltonian. In Appendix B, we derive the explicit form of the electron hopping integrals via ligands as function of an angle characterizing the trigonal distortion of octahedra. In Appendix C, we present the general form of the exchange Hamiltonian including the terms with the Hund's rule coupling, going beyond the symmetric SU(4) form.

\section{Model}

\begin{figure}[t] \centering
\includegraphics[width=0.95\columnwidth]{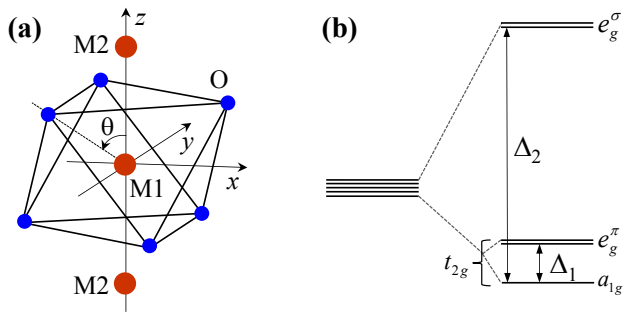}
\caption{(Color online) (a) Magnetic atom (M) surrounded by
trigonally distorted oxygen (O) octahedron in transition metal
compounds with face-sharing octahedra. The global trigonal
coordinate is shown. Trigonal distortion is determined by the
angle $\theta$; the value $\cos\theta_0=1/\sqrt{3}$ corresponds to
undistorted octahedron. Magnetic atoms form a
quasi-one-dimensional chain directed along the $z$ axis. (b)
Crystal field splitting of $d$ orbitals of the magnetic atom. The
splitting of $t_{2g}$ levels ($\Delta_1$) is due to both the
trigonal distortions of oxygen octahedra and contribution from
neighboring M atoms to the crystal field. The sign of $\Delta_1$
can be different depending on the type of
distortions.}\label{FigStruct}
\end{figure}

Let us suppose that we have a linear chain of 3$d$ magnetic ions. Each of them is located at the center of an octahedron of anions with face-sharing geometry. In contrast to the case of corner-sharing octahedra, where the $z$ direction is usually chosen along the fourfold symmetry axis connecting the transition metal ion with one of the apexes of the ligand octahedron (tetragonal coordinate system), here it is convenient to choose trigonal system with the $z$ axis along the chain and the $x$ and $y$ axes in the plane perpendicular to the chain (see Fig.~\ref{FigStruct}a). In such geometry, two nearest-neighbor ions, M1 and M2, are non-equivalent: a pair ligand triangles surrounding one metal ion can be considered as rotated by 180$^{\circ}$ with respect to that surrounding another ion.

To formulate a minimal model for the chain, we start from the well-known Hamiltonian in the second quantization that corresponds
to a general problem of interacting electrons
 \begin{eqnarray} \label{Hubbard}
H&=&\sum_{ij}\sum_{\gamma\gamma'\sigma}t_{ij}^{\gamma\gamma'}
c^{\dag}_{i\gamma\sigma}c_{j\gamma'\sigma}+\\
&&\frac{1}{2}\sum_{i}\!\!\sum_{\gamma\beta\gamma'\beta'}
\!\sum_{\sigma\sigma'}U_{\gamma\beta;\gamma'\beta'}
c^{\dag}_{i\gamma\sigma}c^{\dag}_{i\beta\sigma'}
c_{i\beta'\sigma'}c_{i\gamma'\sigma}\,.\nonumber
\end{eqnarray}
Here, $i$ and $j$ denote lattice sites, where the magnetic ion is located, $\gamma$, $\gamma'$, $\beta$, $\beta'$ run over the active orbitals on each site, $\sigma$, $\sigma'$ denote spin up or spin down and $c_{i\gamma\sigma}$, ($c^{\dag}_{i\gamma\sigma}$) are the annihilation (creation) operators for an electron at site $i$ with the quantum numbers $\gamma$ and $\sigma$. The first term describes the kinetic energy and the second one corresponds to the on-site Coulomb repulsion, where
\begin{equation}\nonumber
U_{\gamma\beta;\gamma'\beta'}=\!
\iint\!d\bm{r}d\bm{r}'\phi^{\star}_{\gamma}(\bm{r})\phi^{\star}_
{\beta}(\bm{r}')V(\bm{r},\bm{r}')
\phi_{\gamma'}(\bm{r})\phi_{\beta'}(\bm{r}')\,.
\end{equation}
Here, $\phi(\bm{r})$ are one-particle wave functions and $V(\bm{r},\bm{r}')$ describes the interparticle interactions. The crystal field felt by the magnetic ions has an important component of cubic $O_h$ symmetry due to octahedra of anions. It splits  the one-electron $d$ levels into a triply degenerate level ($t_{2g}$) and a doubly degenerate level ($e_g$). In the case of the face-sharing octahedra, actual symmetry is usually lower than $O_h$ due to, e.g., axial order of the metal ions, which in such a geometry often form chains, dimers, trimers, etc. This type of low-dimensional packing in its turn results in drastic distortions of the ligand octahedra by itself so that octahedra appear to be trigonally distorted (elongation or compression along the vertical $z$ direction in Fig.~\ref{FigStruct}a). Such local distortions of $D_{3d}$ symmetry lead to splitting of $t_{2g}$ orbitals into an $a_{1g}$ singlet and $e_g^{\pi}$ doublet; the original $e_g$ ($e_g^{\sigma}$) doublet by that remains unsplit (see  below Fig.~\ref{FigStruct}b).

The model treatment will be performed separately for two situations, when $e_g$ and $t_{2g}$ orbitals are active, taking into account trigonal distortions.

\section{$e_g$ levels \label{u-eg}}

We first consider the case of one hole (electron) at the degenerate $e_g$ level, which corresponds e.g. to the orbital filling of
Cu$^{2+}$ ions in CsCuCl$_3$. It has been established (see, e.g. Ref.~\onlinecite{AbragamBleaney1970}) that both the trigonal field
and the spin-orbit coupling do not split the $e_g$ levels.

In the case of ideal MO$_6$ octahedra, one may use the trigonal coordinate system. The $e_g$ doublet for two neighboring magnetic ions along the chain can be written as~\cite{BatesJPhC1971}
\begin{eqnarray}
\mid d_1\rangle &=&\frac{1}{\sqrt{3}}\mid x^2-y^2\rangle-
\sqrt{\frac{2}{3}}\mid xz\rangle\,,\nonumber\\
\mid e_1\rangle &=&-\frac{1}{\sqrt3}\mid xy\rangle
-\sqrt{\frac{2}{3}}\mid yz\rangle\label{e_g_d1}
\end{eqnarray}
for an ion M1, and
\begin{eqnarray}\mid
d_2\rangle &=&\frac{1}{\sqrt{3}}\mid x^2-y^2\rangle
+\sqrt{\frac{2}{3}}\mid xz\rangle\,,\nonumber\\
\mid e_2\rangle &=&-\frac{1}{\sqrt3}\mid xy\rangle
+\sqrt{\frac{2}{3}}\mid yz\rangle\label{e_g_e2} \end{eqnarray} for
the nearest-neighbor ion M2 (the corresponding structure is
illustrated in Fig.~\ref{FigStruct}a).

We start from the two-band 1D Hubbard Hamiltonian of the form of Eq.~\eqref{Hubbard}, where orbital indices $\gamma$ take the values $d_1$, $e_1$ for the M1 sites (sites with e.g. odd $i$) or $d_2$, $e_2$ for the M2 sites (sites with even $i$).  We restrict ourselves by the consideration of the nearest neighbor hopping amplitudes along the chain, $t_{\gamma\gamma'}\equiv t_{ii+1}^{\gamma\gamma'}$. These hopping amplitudes have two contributions, which, for this particular geometry, could be of the same order of magnitude; direct
hopping between two magnetic ions along the chain, $t^{d-d}_{\gamma\gamma'}$, and the indirect hopping via neighboring anions, $t^{viaA}_{\gamma\gamma'}$. We consider both these situations separately.

We begin by calculating the direct hopping terms. We choose the $z$ direction (trigonal axis) parallel to the chain. In this situation, the only monzero $d$\,--\,$d$ Slater--Koster parameters~\cite{SlaterKosterPR1954} are
\begin{eqnarray} t_{xy,xy}&=&t_{x^2-y^2,x^2-y^2}=V_{dd\delta}\,,\nonumber\\
t_{yz,yz}&=&t_{xz,xz}=V_{dd\pi}\,.
\end{eqnarray}
Therefore, we have for the direct case
\begin{equation} \label{direct1}
t^{d-d}\equiv t^{d-d}_{d_2,d_1}=t^{d-d}_{e_2,e_1}=\frac{1}{3}V_{dd\delta}-\frac{2}{3}V_{dd\pi}\,,
\end{equation}
and
\begin{equation} \label{direct2}
t^{d-d}_{e_2,d_1}=t^{d-d}_{d_2,e_1}=0\,.
\end{equation}

The calculation of effective hoppings via ligands $t^{viaA}_{\gamma\gamma'}$ is more complicated. The direct derivation is performed in Appendix B. Here, we only show that $t^{viaA}_{\gamma\gamma'}\propto\delta_{\gamma\gamma'}$ using simple considerations. Assume that we know hopping integrals along a superexchange path between two neighboring cations involving an anion (A1) located at one of the apexes of the octahedron. In general, we have three nonzero hopping integrals $t^{viaA1}_{d_2,d_1}=t_1$, $t^{viaA1}_{e_2,e_1}=t_2$, and
$t^{viaA1}_{d_2,e_1 }=t^{viaA1}_{e_2,d_1}=t_3$ between M1 and M2 ions. Then, the hopping integrals for other two superexchange paths (via A2 and A3) could be found by rotating the $xy$ plane by $\pm\frac{2\pi}{3}$ about the trigonal axis. Denoting by primes the axis in the coordinate system rotated by $\frac{2\pi}{3}$, $(x',y',z')$, $z'=z$, we can write taking into account that $|xy\rangle\propto xy/r^2$ and $|x^2-y^2\rangle\propto(x^2-y^2)/(2r^2)$
\begin{eqnarray}
|d_i\rangle &=&|d'_i\rangle \cos\frac{2\pi}{3}-
|e'_i\rangle \sin\frac{2\pi}{3}\,,\nonumber\\
|e_i\rangle &=&|d'_i\rangle \sin\frac{2\pi}{3}+
|e'_i\rangle\cos\frac{2\pi}{3}\,,
\end{eqnarray}
where $i=1,2$. Therefore
\begin{eqnarray}
|d'_i\rangle &=& |d_i\rangle \cos\frac{2\pi}{3}+
|e_i\rangle \sin\frac{2\pi}{3}\,,\nonumber\\
|e'_i\rangle &=&-|d_i\rangle \sin\frac{2\pi}{3}+
|e_i\rangle\cos\frac{2\pi}{3}\,.\label{de2}
\end{eqnarray}

After the rotation the path M1--A2--M2 becomes the path M1--A1--M2. Thus, we can express the hopping integrals via A2, $t^{viaA2}_{d_2,d_1}=t'_1$, $t^{viaA2}_{e_2,e_1}=t'_2$, and $t^{viaA2}_{d_2,e_1}=t^{viaA2}_{e_2,d_1}=t'_3$ in terms of those via A1 according to $t^{viaA2}_{\mu\nu}=t^{viaA1}_{\mu'\nu'}$. Using Eq.~\eqref{de2}, we obtain
\begin{eqnarray}
t'_1&=&t_1\cos^2\frac{2\pi}{3}+t_2\sin^2\frac{2\pi}{3}+
t_3\sin\frac{4\pi}{3}\,,\nonumber\\
t'_2&=&t_1\sin^2\frac{2\pi}{3}+t_2\cos^2\frac{2\pi}{3}
-t_3\sin\frac{4\pi}{3}\,,\\
t'_3&=&\frac{t_2-t_1}{2}\sin\frac{4\pi}{3}+t_3\cos\frac{4\pi}{3}\,.\nonumber
\end{eqnarray}
The hopping integrals via A3 are found by substituting $\frac{2\pi}{3}$ for $-\frac{2\pi}{3}$:
\begin{eqnarray}
t''_1 &=&t_1\cos^2\frac{2\pi}{3}+t_2\sin^2\frac{2\pi}{3}
-t_3\sin\frac{4\pi}{3}\,,\nonumber\\
t''_2 &=&t_1\sin^2\frac{2\pi}{3}+t_2\cos^2\frac{2\pi}{3}
+t_3\sin\frac{4\pi}{3}\,,\\
t''_3 &=&\frac{t_1-t_2}{2}\sin\frac{4\pi}{3}+
t_3\cos\frac{4\pi}{3}\,. \nonumber
\end{eqnarray}
The total hopping integrals are $t_i^{viaA}=t_i+t'_i+t''_i$ ($i=1,2,3$). Performing the summation over all three paths, we obtain
\begin{eqnarray}
t^{viaA}_{\gamma\gamma'}=t_0\delta_{\gamma\gamma'}\,,\;\;t_0=\frac{3}{2}(t_1+t_2)\,.
\label{t_viaO}
\end{eqnarray}
The value of $t_0$ as a function of the $p$\,--\,$d$ Slater--Koster parameters $V_{pd\sigma}$ and $V_{pd\pi}$, and the $p$\,--\,$d$ charge transfer energy $\Delta$ is calculated in Appendix B.
Here, we see that the situation is again similar to the direct exchange, for which we have equal hopping integrals between the same orbitals, and hopping between different orbitals is absent. This is a rather general result based only on the existence of the threefold trigonal axis and it does not depend on the specific features of the superexchange paths. Therefore, the results  \eqref{direct1}, \eqref{direct2}, \eqref{t_viaO} show that the parameters for the hopping part of the Hamiltonian are $t_{d_2,d_1}=t_{e_2,e_1}=t$ and $t_{e_2,d_1}=t_{d_2,e_1}=0$ with $t=t^{d-d}+t_0$.

For the Coulomb part of Hamiltonian~\eqref{Hubbard}, we can use the standard parametrization: the on-site Coulomb (Hubbard) repulsion on the same orbital $U_{ee,ee} = U_{dd,dd} = U$, and that on different orbitals $U_{de, de} = U' = U-2J$. Here $J$ is
the Hund's rule coupling constant. Note here that the latter relationship is valid only for the unscreened Coulomb potential and can be violated in real transition metal compounds since $U$ is usually screened more by surrounding ligands than the purely
intra-atomic parameter $J$.~\cite{KK-JETP} In the general case, other Slater integrals, not only $U$ and $J$, may
enter~\cite{Khomskii_book2014}; we use below this, the so called
Kanamori parametrization, which in most cases is sufficient.

Assuming that $t \ll (U,J)$, we can change over to an effective Hamiltonian that acts on the subspace of functions with singly occupied sites. The calculation is standard (see, e.g., Refs.~\onlinecite{KK-JETP,KugKhoUFN82}). In the first approximation $(J=0)$, the result is the symmetrical SU(4) model
\begin{equation} H_{eff}=\frac{t^2}{U}\sum_{\langle i,j \rangle }
(\frac{1}{2}+2\bm{s}_i\bm{s}_j)
(\frac{1}{2}+2\bm{\tau}_i\bm{\tau}_j)\, ,\label{KK_model1}
 \end{equation}
where $\bm{s}_i$ is the spin operator of $e_g$  electron at site $i$ defined as $\bm{s}_i=\frac{1}{2}\sum_{\gamma\alpha\beta}c^{\dag}_
{i\gamma\alpha}\bm{\sigma}_{\alpha\beta}c_{i\gamma\beta}$ and $\bm{\tau}_i$ is the pseudospin operator for the orbital degree of
freedom at site $i$ defined as $\bm{\tau}_i=\frac{1}{2}\sum_{\alpha\gamma\gamma'}c^{\dag}_{i\gamma\alpha}
\bm{\sigma}_{\gamma\gamma'}c_{i\gamma'\alpha}$ ($\bm{\sigma}$ are the Pauli matrices). Notice that the same $\bm{\tau}$ operators corresponds to different orbitals at the neighboring sites (since the neighboring face-sharing anion octahedra are rotated with respect to each other). A more general form of the spin-orbital Hamiltonian with the finite Hund's rule coupling $J$ is presented in Appendix C.

Thus, the transition metal compounds with face-sharing octahedra could provide the closest realization of the high-symmetry spin--orbital model. The leading term of the exchange $\sim t^2/U$ has the high SU(4) symmetry, but the terms of higher order containing the Hund's rule coupling constant would have a more complicated form, see Appendix C. The ground state of this general Hamiltonian including terms $\sim J/U$, in the the mean-field approximation is well known to be ferromagnetic in spin and antiferromagnetic in pseudospin.~\cite{KK-JETP,KugKhoUFN82} In general, however, quantum effects related to the SU(4) symmetry may favor other types of states, and the total resulting type of the ground state requires a special analysis.

The value of the effective electron--electron hopping $t$ depends on the details of the crystal structure, in particular, on the
M1--O--M2 angle. Note that at some values of this angle, the contribution of the M1--O--M2 exchange via oxygens can vanish (see
Appendix B), and such case should be treated separately.

\section{$t_{2g}$ levels \label{u-t2g}}

There are many materials, which have the orbital filling corresponding to the present case. These are not only well-known V$_2$O$_3$ and BaVS$_3$, but also many other $3d$ and especially $4d$ and $5d$ transition metal compounds, such as Ba$_4$Ru$_3$O$_{10}$ and BaRuO$_3$. As was mentioned above, even in the case of ideal MO$_6$ octahedra, there exists the trigonal symmetry, which is
inherent to face-sharing geometry.

The trigonal crystal field acts on the triplet $t_{2g}$ level further splitting it into a doublet $(e_g^{\pi})$ and a singlet
$(a_{1g})$. The corresponding part of the  Hamiltonian due to a trigonal field  can be written as
 \begin{equation}
H_t=\delta(L_z^2-\frac{2}{3}I)\,,
 \end{equation}
where $I$ is the unit operator, $L_z$ is the angular  momentum operator in the basis of trigonal axes, and parameter $\delta$ can be positive or negative. We now analyze the sign of the possible contributions to $\delta$. The trigonal field due to a distortion of the octahedra can have both signs, positive for an elongation and negative for a compression of the octahedra along the trigonal axis. The trigonal field due to the neighboring magnetic cations forming 1D structures is always positive ($\delta >0$). The singlet is the lowest energy state for $\delta > 0 $ and the doublet for $\delta<0$.

In the trigonal coordinate system, we have the $a_{1g}$ singlet,
 \begin{equation} |a_1\rangle =
|3z^2-r^2\rangle\,,\label{a1_singlet}
 \end{equation}
and a doublet $e_g^{\pi}$,
 \begin{eqnarray} |b_1\rangle
&=&-\frac{2}{\sqrt{6}}|xy\rangle
+\frac{1}{\sqrt{3}}|yz\rangle\,,\nonumber\\
|c_1\rangle &=&\frac{2}{\sqrt{6}}|x^2-y^2\rangle
+\frac{1}{\sqrt{3}}|xz\rangle\,,\label{t2g_b1}
\end{eqnarray}
for an ion M1, and the same singlet
\begin{equation}
|a_2\rangle =|3z^2-r^2\rangle\,,\label{a2_singlet}
\end{equation}
and a doublet,
\begin{eqnarray}
|b_2\rangle &=&-\frac{2}{\sqrt{6}}|xy\rangle
-\frac{1}{\sqrt{3}}|yz\rangle\,,\nonumber\\
|c_2\rangle &=&\frac{2}{\sqrt{6}}|x^2-y^2\rangle
-\frac{1}{\sqrt{3}}|xz\rangle\,,\label{t2g_c2}
\end{eqnarray}
for the nearest neighbor ion M2.

It has to be mentioned that these expressions for the wave
functions [and Eqs.~\eqref{e_g_d1}\,--\,\eqref{e_g_e2}] are given
for the case of the ideal MO$_6$ octahedra, where M--O--M angle is
about 70.5$^{\circ}$. The trigonal distortions will mix
$e_g^{\pi}$ and $e_g^{\sigma}$ orbitals. More detailed
calculations, which take into account such modification of the
wave function due to trigonal distortions are presented in
Appendix~\ref{td}. This mixing, however, only changes some
numerical coefficients and does not change the main conclusion
that there exist only equal diagonal hoppings, the hopping between
different orbitals being zero -- the conditions important for
getting SU(4) model~\eqref{KK_model1}.

Here, we consider the electronic configuration as shown in Fig.~\ref{FigStruct}b: the $a_{1g}$ level has energy lower than that for the $e_g^{\pi}$ level. The conditions for the existence of such a configuration are discussed in Appendix A. Further on, we assume that the $a_{1g}$ level is fully occupied, there is one electron at the doubly degenerate $e_g^{\pi}$ level, and the upper $e_g^{\sigma}$ levels are empty. In this case, we can use 1D two-band Hubbard Hamiltonian in the form of Eq.~\eqref{Hubbard}, but now orbital indices $\gamma$ take the values $b_1$, $c_1$ for the M1 sites (odd $i$) or $b_2$, $c_2$ for the M2 sites (even $i$). The hopping amplitudes $t_{\gamma\gamma'}$ are the sum of the direct $d$\,--\,$d$ and indirect (via ligands) hopping amplitudes. Note that our analysis is relevant also for the case of negative but large in absolute value $\Delta_1$, when the empty $a_{1g}$ level lies far above the $e_g^{\pi}$ level with one electron.

For the direct $d$\,--\,$d$ hopping, we have now
\begin{eqnarray}
t^{d-d}\equiv t^{d-d}_{b_2,b_1}&=&t^{d-d}_{c_2,c_1} =\frac 23 V_{dd\delta}-\frac 13 V_{dd\pi}\, ,\label{tdd}\\
t^{d-d}_{b_2,c_1}&=&t^{d-d}_{c_2,b_1} =0\,.\nonumber
\end{eqnarray}
To find the relations for the hopping via ligands, we can use the consideration similar to that used in the previous Section, but with the replacement
\begin{equation}
|d_{i}\rangle\to|b_{i}\rangle\,,\;\;|e_{i}\rangle\to|c_{i}\rangle\,,\;\;i=1,\,2\,.
\end{equation}
Repeating after this substitution all calculations as described above, we obtain that the hopping amplitudes have again the symmetric form
\begin{eqnarray}
t_{\gamma\gamma'}=(t^{d-d}+t_0)\delta_{\gamma\gamma'}\,,
\end{eqnarray}
where direct hopping amplitude $t^{d-d}$ is given by Eq.~\eqref{tdd}, while the hopping amplitude via ligands, $t_0$, is obtained in Appendix B [Eq.~\eqref{t_egpi}].

Thus, the same arguments as those presented in the previous
section show that for one electron (or hole) at $e_g^{\pi}$ levels
(neglecting the contribution of $a_{1g}$ states), the effective
spin--orbital Hamiltonian for a chain of face-sharing octahedra
would have the same form of Eq.~\eqref{KK_model1} as for ``real"
$e_g$ orbitals, including the SU(4) part, Eq.~\eqref{KK_model1},
and if necessary the extra terms~\cite{KK-JETP,KugKhoUFN82} $\sim J/U$ (see Appendix C).  This form of the effective
Hamiltonian is, in fact, a consequence of the lattice symmetry:
$e_g^{\pi}$ and $e_g^{\sigma}$ are similar representations of the
same point group. Moreover, taking into account trigonal
distortions of the metal--ligand octahedra and the Coulomb
interaction between cations in the chain does not change the
symmetry of the Hamiltonian (see Appendix~\ref{td} below).

\section{Conclusions}\label{concl}

In the present paper, we considered the effective spin--orbital exchange for the ``third case'' (as compared to the first two well-known cases of MO$_6$ octahedra with common corner and common edge), namely, the case of local geometry with face-sharing MO$_6$ octahedra. The trigonal distortions are inherent to such systems. They determine the symmetry of the problem, splitting the $t_{2g}$ levels to those with $a_{1g}$ and $e_{g}^{\pi}$ orbitals and reduce it to appropriate spin--orbital model with pseudospin-1/2.
We show that resulting effective spin--orbital Hamiltonian in this situation is a well known symmetric Kugel--Khomskii model,
Eq.~\eqref{KK_symmetric}, or, in a more complete form, Eq.~(C1), both for the $e_g^{\sigma}$ and $e_{g}^{\pi}$ orbitals. The leading terms of the model have the SU(4) symmetry. In that sense, the situation with face-sharing geometry is very different from the usually considered cases of MO$_6$ octahedra with a common corner (M1--O--M2 angle $\sim 180^{\circ}$) and with a common edge (M1--O--M2 angle $\sim 90^{\circ}$).

This result is important in several  respects. First of all, it points out a class of real physical systems, for which the spin--orbital model of SU(4) symmetry can be applied. This opens the possibility to experimentally check some nontrivial predictions of this model, such as strong spin--orbital entanglement and crucial role of quantum effects. Second, it is instructive to compare the general tendencies existing for three typical geometries: those of MO$_6$ octahedra with a common corner (one common oxygen for two neighboring MO$_6$ octahedra), common edge (two common oxygens), and a common face (three common oxygens). The general conclusions in the better known first and second cases are rather different. For the common-corner geometry, the typical well-known rule is that the ferro-orbital ordering
gives antiferromagnetic spin alignment, and {\it vice versa}.~\cite{Gooden_book,Khomskii_book2014,KK-JETP,KugKhoUFN82} However, this is not true for the case of common edges, with $\sim 90^{\circ}$ M1--O--M2 bonds: in this situation, often one has ferromagnetic
spin ordering irrespective of orbital occupation.~\cite{MostKhomPRL2002,StrKhomPRB2008pyroxenes} In that sense, the situation with face-sharing octahedra leading, e.g., to Hamiltonian~\eqref{KK_model1} is more similar to that with a common corner than to the situation with a common edge: ferro-spins coexist with antiferro-orbitals and {\it vice versa}. On the other hand, as stressed in Appendix B, for the superexchange via ligands (but not for direct $d-d$ hopping!) the leading terms in the exchange $\sim t^2/U \sim [t_{pd}^2/\Delta]^2/U$ can drop out for certain values of the M1--O--M2 angle, similar to the case of common-edge geometry. Thus, the systems with face-sharing geometry represent a class of their own, and they have to be considered as such. Our treatment is focused on the specific features related to such geometry, and the resulting picture turns out to be quite interesting.

Turning to real systems, several factors not considered in the present paper may become important, which could decrease the symmetry of the resulting model. One is the electron-lattice (Jahn--Teller) interaction, which, in principle, could lead to orbital ordering independent of the spin one; in systems like CsCuCl$_3$, for example, it could result in helicoidal  superstructures (see Ref.~\onlinecite{Bersuker_book} and references therein). The second one, considered in detail in the Appendix B, is the strong trigonal distortion of MO$_6$ octahedra, which for particular situations can strongly reduce the M--O--M contribution to the superexchange, so that for certain M--O--M angle, only the direct $d-d$ contribution remains. In this case, one may need to take into account higher-order terms $\sim J/U$ in the superexchange Hamiltonian. These terms, written down in the general expression (C1) presented in Appendix~C, have less symmetric form in orbital variables $\tau$; i.e., they can also violate the SU(4) symmetry. Nevertheless, pronounced quantum effects typical of the SU(4) model with its intrinsic strong spin-orbital entanglement can still can still be dominant and determine the type of the ground state of the system. However, even if the type of the ground state at $T=0$ would be determined by these symmetry-breaking terms (with the energy scale $J' \sim \frac{t^2}{U}\frac{J}{U}$, which is are typically about 10\% of the main SU(4) term of the order of  $\frac{t^2}{U}$), there would exist a broad temperature range  $J'\lesssim T\lesssim\frac{t^2}{U}$, in which the behavior would be determined by the SU(4) physics. However, the situation taking place in each particular real system requires a special treatment.

As far as real materials are concerned, one more issue is worth discussing. Whereas the situation with common corner and common edge geometry is met in all cases, 3D, 2D, and 1D, common-face geometry in this sense is more ``choosy": it is typical  for one-dimensional systems  (CsCuCl$_3$, BaCoO$_3$), and often such face-sharing octahedra exist for just dimers or linear trimers (e.g. in BaIrO$_3$). We are not aware of any real substances with the 2D or 3D face-sharing geometry, although we cannot exclude such cases in principle. As to the exchange in such hypothetical situations, we can give some arguments that in this case for real $e_g$ systems the resulting Hamiltonian would also be in a first approximation SU(4)-symmetric, but for the $t_{2g}$ levels it would not be the case, because the choice of relevant $a_{1g}$ and $e_g^{\pi}$ orbital would depend on the direction and be different for different nearest neighbors. The 1D systems, however, should not necessarily involve straight chains; there may be zigzag or even spiral chains. In all such cases, the SU(4) physics would be preserved for $e_g$ electrons to a first approximation. In some sense, it might be even advantageous, because such 1D model is exactly soluble -- although it would be very interesting (if at all possible) to have similar 2D or 3D systems.

\begin{acknowledgments}
This work is supported by the Russian Foundation for Basic Research (projects 14-02-00276-a, 14-02-0058-a, 13-02-00050-a, 13-02-00909-a, and 13-02-00374-a), by the Russian Science Support Foundation, by the Ministry of Education and Science of Russia (grant MK 3443.2013.2), by the Ural Branch of Russian Academy of Sciences, by the German projects DFG GR 1484/2-1 and FOR 1346, and by K\"oln University via the German Excellence Initiative.
\end{acknowledgments}

\appendix \section{Face-sharing octahedra with trigonal
distortions \label{td}}

Let us now  consider a more general case, namely, that with the
crystal field of trigonal symmetry corresponding to the stretching
or compression of the chain of face-sharing octahedra. In the main
text, we considered exchange interaction for $e_g$ and $t_{2g}$
orbitals taking for the corresponding wave functions those of pure
$e_g$ and $t_{2g}$ orbitals for cubic symmetry. However, trigonal
distortion can modify these wave functions, leading, in
particular, to a mixing of $e_g^{\sigma}$ and $e_g^{\pi}$ orbitals.
In this Appendix, we consider these effects; as a result, we find
that their inclusion does not qualitatively modify our main
conclusions, and can lead only to some change in certain numerical
coefficients.

An elementary building block of transition metal compounds with
face-sharing octahedra is shown in Fig.~\ref{FigStruct}a. Magnetic
atoms form a quasi-one-dimensional chain directed along the $z$
axis. Each magnetic atom is surrounded by the distorted oxygen
octahedron. Distortions are described by a single parameter
$\theta$, which is the angle between the $z$ axis and the line
connecting M and O atoms (see Fig.~\ref{FigStruct}a). For
undistorted octahedron, we have
$\theta=\theta_0=\arccos(1/\sqrt{3})$. The crystal field splits
fivefold-degenerate $d$ electron levels of the transition metal
atom into two doubly degenerate $e_{g}^{\sigma}$, $e_{g}^{\pi}$
levels, and $a_{1g}$ level, as it is shown in
Fig.~\ref{FigStruct}b. The energy difference $\Delta_1$ between
$e_{g}^{\pi}$ and $a_{1g}$ levels can be positive or negative
depending on the type of trigonal distortions. Stretching of
oxygen octahedron ($\theta<\theta_0$) increases the energy of the
$a_{1g}$ level with respect to $e_{g}^{\pi}$ one, leading to
$\Delta_1<0$. However, the contribution to the crystal field from
a neighboring magnetic cations acts in opposite direction, and, in
general, we can have $\Delta_1>0$ even for (slightly) stretched
octahedra.

Let us now discuss some details. In the point-charge
approximation, the crystal field potential acting onto a chosen
cation located at point $\mathbf r$ can be represented as a sum of
Coulomb terms
\begin{equation} V(\mathbf r) =
v_0\sum_i\frac{r_0}{|{\mathbf r} - {\mathbf r}_i|}\, ,
\end{equation}
where ${\mathbf r}_i$ are the positions of ligand
ions. For $d$ states, the existence of the threefold symmetry axis
leads to a significant simplification of the expression for the
crystal field, which can be, approximately, written in the following form
 \begin{equation}
V({\mathbf r}) = V_0(r) +v_1(r)\sum_{s=1}^{3} P_2(\cos\theta_s) +
v_2(r)\sum_{s=1}^{3}P_4(\cos\theta_s), \end{equation} where $P_2$
and $P_4$ are the Legendre polynomials, $P_2(x) =
\frac{1}{2}(3x^2- 1)$ and $P_4(x) = \frac{1}{8}(34x^4- 30x^2 +
3)$. Here, we took into account  the symmetry in the arrangement
of two opposite edges of the ligand octahedron and as a result, we
have
 \begin{equation}
\cos\theta_s=\cos\theta'\cos\theta+\sin\theta'\sin\theta\cos
\left(\phi'-\frac{2\pi s}{3}\right),
 \end{equation}
where $\theta'$ and $\phi'$ describe the direction of
$\mathbf{r}$,  that is,
$\mathbf{r}=r\{\sin\theta'\cos\phi',\,\sin\theta'\sin\phi',\,\cos\theta'\}$.

Now, we should find the matrix elements of the crystal field for
the complete set of $d$ functions \begin{eqnarray}
|xy\rangle&=&R_d(r){\sin^2\theta'\sin2\phi'\over 2}\,,\nonumber\\
|xz\rangle&=&R_d(r)\sin\theta'\cos\theta'\cos\phi'\,,\nonumber\\
|yz\rangle&=&R_d(r)\sin\theta'\cos\theta'\sin\phi'\,,\label{d-functions}\\
|x^2 -y^2\rangle&=&R_d(r){\sin^2\theta'\cos2\phi'\over2}\,,\nonumber\\
|2z^2-x^2-y^2\rangle&=&R_d(r){{3\cos^2\theta'\
-1}\over2}\,.\nonumber \end{eqnarray}

Straightforward, but rather cumbersome calculations, lead us to the
following matrix \begin{eqnarray}
\hat{V}_{\alpha\beta}&=&E_0\!\times \nonumber \\
\!\!\!&&\!\!\!\!\!\!\!\!\!\!\!
\!\!\!\!\!\!\!\!\!\!\!\!\!\!\left(\!
\begin{array}{ccccc}
\!\frac{-3a_4-10a_2}{15}\!\!\!&\!0\!\!\!&\!\frac{-b}{2}\!\!\!&\!0\!\!&\!0\\
\!0\!\!\!\!&\!\frac{12a_4+5a_2}{15}\!\!\!&\!0\!\!\!&\!\frac{b}{2}\!\!\!&\!0\\
\!\frac{-b}{2}\!\!\!&\!0\!\!\!&\!\frac{12a_4+5a_2}{15}\!\!\!&0\!\!\!&\!0\\
\!0\!\!\!&\!\frac{b}{2}\!\!\!&0\!\!\!&\!\frac{-3a_4-10a_2}{15}\!\!\!&\!0\\
\!0\!\!\!&\!0\!\!\!&\!0\!\!\!&\!0\!\!\!&\!\frac{-18a_4+10a_2}{15}\!\\
\end{array}
\!\!\!\right)\!,\nonumber
\end{eqnarray}
where $E_0 =10Dq$ is the splitting between $e_g$ and $t_{2g}$ levels, and
\begin{eqnarray}
a_4&=&-{3\over 2}\left({5\over 2}\cos^4\theta -
{15\over 7}\cos^2\theta + {3\over 14}\right)\,,\nonumber\\
a_2&=&{27\over 35}\kappa\left(3\cos^2\theta -1\right)\,,
\label{CF_parameters}\\
b&=&3\sin^3\theta\cos\theta\,.\nonumber
\end{eqnarray}
Here, parameter $\kappa$ is defined as
 \begin{equation}
\kappa
=r_0^2\frac{\int_0^{\infty}r^2R_d^2(r)r^2dr}
{\int_0^{\infty}r^4R_d^2(r)r^2dr}=
r_0^2\frac{\langle r^2\rangle} {\langle r^4\rangle}
=k\left({r_0}\over{a_B}\right)^2\,,
 \end{equation}
where $r_0$ is the cation--ligand distance and $a_B$ is the Bohr radius. A  rough estimate for the factor $k$ can be found by using the hydrogen-like form for the radial part $R_d(r)$ of the wave function in metal ions, $R_d(r)\sim r^{n^*-1}\exp{(-z^*r/a_B)}$, where $n^*$ and $z^*$, are the effective values of the principal quantum number and of the nuclear charge, respectively.~\cite{SlaterPR1930} According to Ref.~\onlinecite{SlaterPR1930}, we have $n^*=3$, $3.7$, and $4$ for 3$d$, 4$d$, and 5$d$ shells, respectively. For $d$ electrons, there is the following simple rule: the charge of all filled shells inside the $d$ shell is subtracted from the nuclear charge and the charge of all $d$ electrons except the given one is multiplied by 0.35 and also subtracted. For example, for Co$^{4+}$ with the nuclear charge $z =27$, we find $z^* = 7.6$. In this
case, we have $k = (z^*)^2/810 \approx 0.07$. More accurate estimates using the linearized muffin-tin orbitals (LMTO) give $k = 0.2-0.3$.

\begin{figure}[t] \centering
\includegraphics[width=0.75\columnwidth]{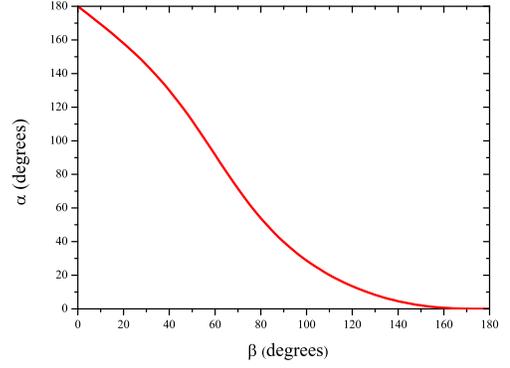}
\caption{(Color online) Angle $\alpha$ versus M1--O--M2 angle
$\beta =\pi -2\theta$; $k = 0.1$, $r_0 = 2$ \AA.}\label{alpha1}
\end{figure}

As a result, we find the wave functions of $e_{g}^{\sigma}$, $e_{g}^{\pi}$, and $a_{1g}$ energy levels, which depend on the trigonal distortions. Choosing the reference frame like shown in Fig.~\ref{FigStruct}a, we obtain for the wave functions expressions having the forms similar to those obtained above for the case of undistorted octahedra. Thus, for $e_g$ levels
($e_g^{\sigma}$ orbitals) we have [cf. Eqs.~\eqref{e_g_d1} and \eqref{e_g_e2}]
\begin{eqnarray}
|d_{1,2}\rangle&=&\sin\!\frac{\alpha}{2}\,|x^2-y^2\rangle
\mp\cos\!\frac{\alpha}{2}\,|xz\rangle\,,\nonumber\\
|e_{1,2}\rangle&=&-\sin\!\frac{\alpha}{2}\,|xy\rangle
\mp\cos\!\frac{\alpha}{2}\,|yz\rangle\,.\label{e_g_d12_gen}
\end{eqnarray}
For $t_{2g}$ orbitals, we have the same $a_{1g}$ singlet, Eqs.~\eqref{a1_singlet} and \eqref{a2_singlet}, and the
$e_g^{\pi}$ doublet  [cf. Eqs.~\eqref{t2g_b1}\,--\,\eqref{t2g_c2}]
\begin{eqnarray}
|b_{1,2}\rangle&=&-\cos\!\frac{\alpha}{2}\,|xy\rangle
\pm\sin\!\frac{\alpha}{2}\,|yz\rangle\,,\nonumber\\
|c_{1,2}\rangle&=&\cos\!\frac{\alpha}{2}\,|x^2-y^2\rangle
\pm\sin\!\frac{\alpha}{2}\,|xz\rangle\,.\label{t2g_c12_gen}
\end{eqnarray} 
The $\mp$ and $\pm$ signs in the above expressions for cation wave functions for neighboring M atoms occur since the oxygen octahedra surrounding neighboring metal atoms are transformed to each other by the $180^{\circ}$ rotation about $z$ axis. Parameter $\alpha$ in
Eqs.~\eqref{e_g_d12_gen} and \eqref{t2g_c12_gen} depends on the
trigonal distortions as well on the contribution to the crystal
field from magnetic atoms. Neglecting the latter effect, we find
\begin{equation} \cos \alpha = \frac{a}{\sqrt{a^2+b^2}}, \,\,\, a
= a_2 + a_4. \label{alpha}
 \end{equation}
For the ideal octahedron, we have
$\alpha=\alpha_0\equiv\pi-2\theta_0 = \arccos(1/3)$. Substituting
this value to Eqs.~\eqref{e_g_d12_gen} and \eqref{t2g_c12_gen},
we arrive at the results of the previous sections. The dependence
of $\alpha$ on the M1--O--M2 angle $\beta=\pi-2\theta$ is
illustrated in Fig.~\ref{alpha1}. Parameter $\alpha$ decreases monotonically when $\beta$ increases, and it changes faster for $\beta$ close to the value $\beta_0=\pi-2\theta_0$ corresponding to the ideal octahedron. This decrease becomes sharper for larger values of $\kappa$.

These results were obtained neglecting the effect of neighboring
metal atoms in the chain. Taking into account the contribution to
the crystal field from these atoms modifies the parameter $a_2$ in
the following manner
 \begin{equation} a_2\to
a_2-\frac{27\kappa}{35}\frac{Z^*}{12\cos^2\theta}\,,
 \end{equation}
where $Z^{*}$ is the effective charge (in units of
$e$) of the metal ion. Note that $Z^{*}$ can be different from
$z^{*}$ mentioned above. Parameters $a_4$ and $b$, as well as the
relations~\eqref{e_g_d12_gen}--\eqref{alpha} remain the same.
Stretching of oxygen octahedra ($\theta<\theta_0$) tends to make
$\alpha<\alpha_0$, while the effect of neighboring metal atoms
acts in the opposite direction. For $\alpha>\alpha_0$, the energy
of the $a_{1g}$ level is lower than that of the $e_{g}^{\pi}$ one (see Fig.~\ref{FigStruct}b), leading to $\Delta_1>0$. Thus, in general, we can have $\Delta_1>0$ even for (slightly) stretched octahedra. Just this situation takes place in BaCoO$_3$ with the chains of face-sharing Co$^{4+}$O$_6$
octahedra.~\cite{YamauraJSStCh1999BaCoO3} Here, Co$^{4+}$ with the
$d^5$ configuration has one hole at the $e_g^{\pi}$ level.

The wave functions \eqref{e_g_d12_gen} and \eqref{t2g_c12_gen} are the generalization of those considered in the Sec.~\ref{u-eg}\,--\,\ref{u-t2g} to the case of arbitrary trigonal distortion characterized by an angle $\alpha$. In other words, these distortions mix the $e_g^{\pi}$ and $e_g^{\sigma}$ wave functions for ideal octahedra MO$_6$ given in Eqs.~\eqref{e_g_d1} and \eqref{e_g_e2} and Eqs.\eqref{t2g_b1}\,--\,\eqref{t2g_c2}.

It is quite straightforward to demonstrate that orbitals \eqref{e_g_d12_gen} and \eqref{t2g_c12_gen} provide the structure of the spin--orbital Hamiltonian of the same form of Eq.~\eqref{KK_model1} at any given $\alpha$ (taking into account both direct and ligand-assistant hoppings). Thus, our main conclusions remain the same even with the $e_g^{\sigma}-e_g^{\pi}$
mixing taken into account.

\section{Electron hopping via ligands \label{hopping}}

Here, we analyze a possible dependence of the hopping integrals between metal ions via ligand ions (let them call oxygens for brevity) on the M-O-M bond angle. In the chain of face sharing
MO$_6$ octahedra, we chose a unit cell consisting of two oxygen triangles forming an octahedron and two metal ions, M1 and M2 (see Fig.~\ref{FigStruct}). Then, the tight-binding Hamiltonian describing the charge transfer between metal ions via oxygen can be written in the following form (spin indices are omitted for simplicity)
\begin{eqnarray}
H_{pd}&=&-\sum_{n\mu A}\left[t_{1\mu;1A}d_{n1\mu}^{\dag}p^{\phantom{\dag}}_{n1A}
+ t_{2\mu;1A}d_{n2\mu}^{\dag}p^{\phantom{\dag}}_{n1A}\right. \nonumber \\
&&\left.+t_{1\mu;2A}d_{n1\mu}^{\dag}p^{\phantom{\dag}}_{n2A}+
t_{2\mu;2A}d_{n2\mu}^{\dag}p^{\phantom{\dag}}_{n-1\,2A} + H.c.\right]\nonumber \\
&&+\Delta\sum_{nA}\left(p_{n1A}^{\dag}p^{\phantom{\dag}}_{n1A}+
p_{n2A}^{\dag}p^{\phantom{\dag}}_{n2A}\right)\,,\label{H_pd}
 \end{eqnarray}
where $n$ enumerates unit cells, $p^{\dag}(p)$ and $d^{\dag}(d)$ are creation (annihilation) operators for $p$ and $d$ electrons, respectively, numbers 1 and 2 correspond to metal ions M1 and M2, respectively, and to the oxygen triangle above each of them, $\mu$ is the set of basis $d$ functions [Eqs.~\eqref{e_g_d12_gen} for $e_g^{\sigma}$ orbitals or Eqs.~\eqref{t2g_c12_gen} for $e_g^{\pi}$ orbitals], and $A = \{s, \eta\}$, where $s=1, 2, 3$ and $\eta = p_x, p_y, p_z$, is a set of subscripts numbering the
atoms in each oxygen triangle and denoting the oxygen $p$ orbitals. For each doublet ($e_g^{\sigma}$ or $e_g^{\pi}$), we
have four $d_{nl\mu}$ ($l=1,2$) operators and eighteen $p_{njA}$ ($j=1,2$) operators, for which we should take into account all possible
electron hoppings.

In the second order of the perturbation theory on $t_{pd}/\Delta$, we can derive a Hamiltonian describing the effective hoppings of electrons between the states of $d$ doublets under study via the oxygen $p$ orbitals. To do this, we first proceed to the momentum representation for electronic operators $d_{kl\mu}=\sum_ne^{-2ikz_0n}d_{nl\mu}/\sqrt{N}$,  $p_{kjA}=\sum_ne^{-2ikz_0n}p_{njA}/\sqrt{N}$, where $N$ is the number of unit cells in the chain and $z_0=2r_0\cos\theta$ is the distance between neighboring M$1$ and M$2$ atoms ($r$ is the M--O distance). Following then the standard procedure, we obtain for the effective Hamiltonian
\begin{equation}\label{H_eff}
H_{eff}=-\sum_{k\mu\nu}\left[t_{\mu\nu}(k)d_{k1\mu}^{\dag}d^{\phantom{\dag}}_{k2\nu}+H.c.\right]\,, \end{equation}
where ($t_{l\mu;jA}$ are assumed to be real)
\begin{equation}\label{tmunu}
t_{\mu\nu}(k)=\frac{1}{\Delta}\sum_{A}\left[t_{1\mu;1A}t_{2\nu;1A}+t_{1\mu;2A}t_{2\nu;2A}e^{-ikz_0}\right].
\end{equation}

According to Ref.~\onlinecite{SlaterKosterPR1954}, the hopping amplitudes $t_{l\mu;jA}$ can be expressed via two Slater--Koster parameters $V_{pd\sigma}$ and $V_{pd\pi}$ and
directing cosines of the radius vector $\mathbf{r}$ connecting the corresponding oxygen and metal ions~\cite{note}. If we choose the reference frame as shown in Fig.~\ref{FigStruct}, the radius vector $\mathbf{r}_{l;js}$ directed from the oxygen atom $s$($=1,2,3$) in the $j$th ($j=1,2$) group of oxygens to the neighboring metal ion $l$($=1,2$) is
\begin{equation}
\mathbf{r}_{l;js}=r_0(-1)^{j}\!\{\sin\theta\cos\varphi_s,
\,\sin\theta\sin\varphi_s,\,(-1)^{l-1}\!\!\cos\theta\},
\end{equation}
where $\varphi_s=2\pi(s-1)/3$. Using these relations, Table~I of Ref.~\onlinecite{SlaterKosterPR1954}, and Eqs.~\eqref{e_g_d12_gen} for $e_g^{\sigma}$ orbitals or Eqs.~\eqref{t2g_c12_gen} for $e_g^{\pi}$ orbitals, we calculate the hopping amplitudes $t_{l\mu;jA}$ as functions of Slater--Koster parameters $V_{pd\sigma}$ and $V_{pd\pi}$, the angle $\theta$, and the parameter $\alpha$ describing the orbital states. Substituting then the obtained $t_{l\mu;jA}$ into Eq.~\eqref{tmunu} and performing the summation, we arrive finally  to the following relation for the effective $d$--$d$ hopping amplitudes
\begin{equation}\label{t0munu}
t_{\mu\nu}(k)=\delta_{\mu\nu}t_0\left(1+e^{-ikz_0}\right)\,.
\end{equation}
This relation is valid both for $e_g^{\sigma}$ and $e_g^{\pi}$ orbitals. For $e_g^{\sigma}$ orbitals, the parameter $t_0$ is
\begin{eqnarray}
t_0&\!\!\!=\!\!\!&-\frac{9}{8}\frac{V_{pd\sigma}^2}{\Delta}\sin^2\!\theta\cos2\theta
\!\left(\!2\cos\theta\cos\frac{\alpha}{2}-\sin\theta\sin\frac{\alpha}{2}\right)^{\!2}\!\!\!+\nonumber\\
&&\frac{3}{2}\frac{V_{pd\pi}^2}{\Delta}\left[\left(\sin\theta\sin\frac{\alpha}{2}
+\cos\theta\cos\frac{\alpha}{2}\right)^2+\right.\nonumber\\
&&\left.\cos2\theta\left(\sin\theta\cos\theta\sin\frac{\alpha}{2}
-\cos2\theta\cos\frac{\alpha}{2}\right)^2\right]+\nonumber\\
&&\frac{3\sqrt{3}}{2}\frac{V_{pd\sigma}V_{pd\pi}}
{\Delta}\sin\theta\sin2\theta\left[\sin^2\theta\cos\theta\sin^2
\frac{\alpha}{2}-\right.\nonumber\\
&&\sin\theta\left(3\cos^2\theta-\sin^2\theta\right)
\sin\frac{\alpha}{2}\cos\frac{\alpha}{2}+\nonumber\\
&&\left.2\cos\theta\cos2\theta\cos^2\frac{\alpha}{2}\right].
\label{t_egsigma}
\end{eqnarray}
In the case of $e_g^{\pi}$ electrons, the hopping integral $t_0$ reads
\begin{eqnarray}
t_0&\!\!\!=\!\!\!&-\frac{9}{8}\frac{V_{pd\sigma}^2}{\Delta}\sin^2\!\theta\cos2\theta
\!\left(\!2\cos\theta\sin\frac{\alpha}{2}-\sin\theta\cos\frac{\alpha}{2}\right)^{\!2}\!\!\!+\nonumber\\
&&\frac{3}{2}\frac{V_{pd\pi}^2}{\Delta}
\left[\left(\sin\theta\cos\frac{\alpha}{2}
+\cos\theta\sin\frac{\alpha}{2}\right)^2+\right.\nonumber\\
&&\left.\cos2\theta\left(\sin\theta\cos\theta\cos\frac{\alpha}{2}
-\cos2\theta\sin\frac{\alpha}{2}\right)^2\right]+\nonumber\\
&&\frac{3\sqrt{3}}{2}\frac{V_{pd\sigma}V_{pd\pi}}
{\Delta}\sin\theta\sin2\theta\left[\sin^2\theta\cos\theta\cos^2
\frac{\alpha}{2}-\right.\nonumber\\
&&\left. \sin\theta\left(3\cos^2\theta-\sin^2\theta\right)
\sin\frac{\alpha}{2}\cos\frac{\alpha}{2}+ \nonumber \right.\\
&&\left.2\cos\theta\cos2\theta\sin^2\frac{\alpha}{2}\right].\label{t_egpi}
\end{eqnarray}

Note, that the effective Hamiltonian~\eqref{H_eff} with hopping amplitudes of the form of Eq.~\eqref{t0munu} is equivalent to the simple tight-binding Hamiltonian of the form
\begin{equation}
H_{eff}=-t_0\sum_{m\mu}\left[d_{m\mu}^{\dag}d^{\phantom{\dag}}_{m+1\mu}+H.c.\right]\,.
\end{equation}
This can be easily checked by using the transformation for electronic operators
\begin{equation}
d^{\phantom{\dag}}_{n1\mu}\to d^{\phantom{\dag}}_{2m\mu},\;\;d^{\phantom{\dag}}_{n2\mu}\to d^{\phantom{\dag}}_{2m+1\mu},\;\;m\in{\cal Z}\,.
\end{equation}
Thus, from viewpoint of the electronic properties, the magnetic sites M$1$ and M$2$ are equivalent to each other even though crystallographically they are different. One should keep in mind, however, that $d$-orbitals wave functions of neighboring magnetic sites are different.

\begin{figure}[t] \centering
\includegraphics[width=0.85\columnwidth]{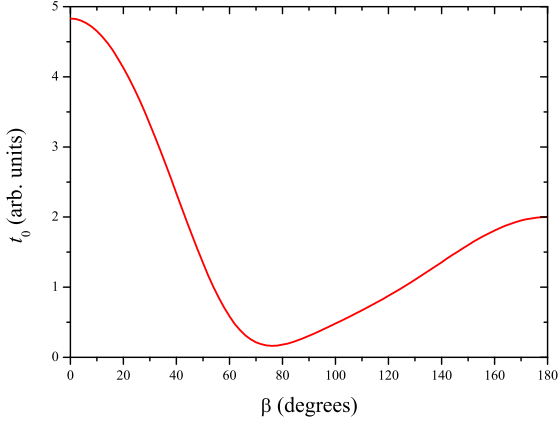}
\caption{(Color online) Hopping integral for $e_g^{\sigma}$
orbitals versus M1--O--M2 angle $\beta =\pi -2\theta$; $k = 0.1$,
$r_0 = 2$ \AA.}\label{hop-sigma} \end{figure}

\begin{figure}[t] \centering
\includegraphics[width=0.85\columnwidth]{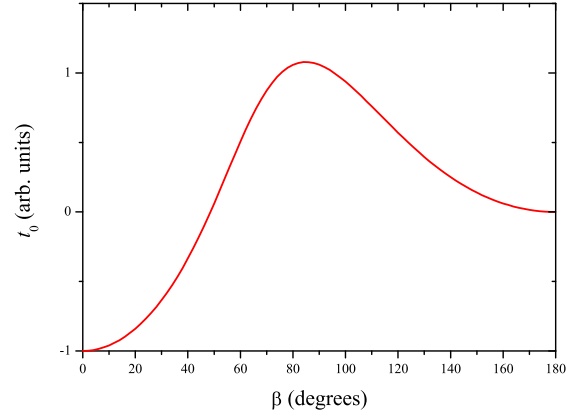}
\caption{(Color online) Hopping integral for $e_g^{\pi}$ orbitals
versus M1--O--M2 angle $\beta =\pi -2\theta$; $k = 0.1$, $r_0 = 2$
\AA.}\label{hop-pi} \end{figure}

The dependence of the hopping integral $t_0$ for
$e_g^{\sigma}$ and $e_g^{\pi}$ orbitals on the M--O--M bond angle $\beta=\pi-2\theta$ is illustrated in
Figs.~\ref{hop-sigma} and \ref{hop-pi}, respectively. For the
ratio $V_{pd\sigma}/V_{pd\pi}$, we took the commonly used
value~\cite{Harrison_book} equal to $2.16$. We see that at some value of $\beta$ the hopping integral
via oxygens either changes sign (for $e_g^{\pi}$ orbitals) or becomes close to zero (for $e_g^{\sigma}$ orbitals). For $e_g^{\sigma}$ orbitals this happens for M--O--M bond angle close to that
characteristic of an undistorted octahedron $\beta_0=\pi-2\arccos(1/\sqrt{3})\cong70.5^{\circ}$, while for $e_g^{\pi}$ orbitals $t_0$ changes the sign at a bit smaller value of $\beta$ (compressed octahedron). The total $d$--$d$ hopping amplitude is $t=t_0+t^{d-d}$. Thus, for $e_g^{\sigma}$ orbitals $t$ is always positive, while for $e_g^{\pi}$ orbitals it can change sign. Usually, the direct $d$--$d$ hopping amplitude $t^{d-d}$ is assumed to be smaller than the characteristic value of the effective hopping via oxygens $t_0\sim V_{pd\sigma}^2/\Delta$. Our calculations show however, that for some M--O--M bond angles the direct hopping becomes dominant. Moreover, for $e_g^{\sigma}$ orbitals this can be the case of the ideal octahedron. When the hopping is suppressed, the higher-order corrections to the SU(4) model, containing the terms $\sim J/U$, which have less symmetric form in orbital $\tau$ variables (see e.g. Refs.~\onlinecite{KK-JETP} and  \onlinecite{KugKhoUFN82}) may have to be included.

\section{General form of the exchange Hamiltonian \label{J/U}}

Let us now present the full form of the exchange Hamiltonian, including terms containing the Hund's rule coupling constant $J$. These terms appear when we consider not the virtual hopping between occupied orbitals, but the hopping to an empty orbital of the neighbor, with the consecutive effect of Hund's coupling. The derivation of these terms is straightforward~\cite{KK-JETP,KugKhoUFN82}, although a bit cumbersome. In our case of face-sharing octahedra, the resulting spin--orbital Hamiltonian has the form
\begin{eqnarray}
&&H_{eff}=\frac{t^2}{U}\sum_{\langle i,j \rangle}
\left\{ (\frac{1}{2}+2\bm{s}_i\bm{s}_j)(\frac{1}{2}+
2\bm{\tau}_i\bm{\tau}_j)+\right.\label{KK_modelJ} \\
&&\frac{JU}{U^2-J^2}[2(\bm{\tau}_i\bm{\tau}_j
-\tau^z_i\tau^z_j)-(\frac{1}{2}+
2\bm{s}_i\bm{s}_j)(\frac{1}{2}-2\tau^z_i\tau^z_j)]
+\nonumber \\
&&\left.\frac{J^2}{U^2-J^2}[(2\tau^z_i\tau^z_j -\frac{1}{2})
+(1+2\bm{s}_i\bm{s}_j) (\bm{\tau}_i\bm{\tau}_j-\tau^z_i\tau^z_j)]
\right\}\,.\nonumber
\end{eqnarray}

We see indeed that, whereas the leading term in the full Hamiltonian \eqref{KK_modelJ} has the SU(4) form, its symmetry is broken by the terms of higher order in $J/U$. Nevertheless, for many realistic situations these terms give only a small correction to the main term, which could in principle be smaller that the results of quantum fluctuations in the SU(4) model. And, as mentioned above, even if these symmetry-breaking terms would determine the type of the ordering in the ground state at $T=0$, there would exist a broad temperature range $(t^2/U)(J/U)\lesssim T\lesssim t^2/U$, in which the properties of the system would be determined by the first, SU(4), part of the Hamiltonian.

\end{document}